\documentclass[aps,prd,twocolumn,showpacs,superscriptaddress,amsmath]{revtex4-1}

\usepackage[dvipdfm]{graphicx}
\usepackage[dvipdfm]{hyperref}

\newcommand{\unit}[1]{~\mathrm{#1}}
\newcommand{\To}{T_{\oplus}}
\newcommand{\omegao}{\omega_{\oplus}}

\newcommand{\rot}{{\rm rot}}
\newcommand{\mpr}{m^{\prime}}
\newcommand{\CCmm}{C^{C}_{m \mpr}}
\newcommand{\CSmm}{C^{S}_{m \mpr}}
\newcommand{\SCmm}{S^{C}_{m \mpr}}
\newcommand{\SSmm}{S^{S}_{m \mpr}}
\def\voc{\mathrel{\rlap{\lower0pt\hbox{\hskip1pt{$c$}}}\raise3pt\hbox{$\neg$}}}
\def\sk#1#2#3{#1^{(#2)}_{#3}}
\def\cf{{\sk{(\voc_F^{(d)})}{0E}{njm}}}
\def\cfdnjm#1#2{{\sk{(\voc_F^{(#1)})}{0E}{#2}}}
\renewcommand{\Re}{{\rm Re}}
\renewcommand{\Im}{{\rm Im}}
\newcommand{\Recfdnjm}[2]{\Re\big[\cfdnjm{#1}{#2}\big]}
\newcommand{\Imcfdnjm}[2]{\Im\big[\cfdnjm{#1}{#2}\big]}
\def\rf#1{(\ref{#1})}
\def\CC#1{C^{C}_{#1}}
\def\CS#1{C^{S}_{#1}}
\def\SC#1{S^{C}_{#1}}
\def\SS#1{S^{S}_{#1}}

\begin{document}

\title{Optical-Cavity Limits on Higher-Order Lorentz Violation}

\author{Yuta Michimura}
%\email{michimura@granite.phys.s.u-tokyo.ac.jp}
\affiliation{Department of Physics, University of Tokyo, Bunkyo, Tokyo 113-0033, Japan}
\author{Matthew Mewes}
%\email{mmewes1@swarthmore.edu}
\affiliation{Department of Physics and Astronomy, Swarthmore College, Swarthmore, Pennsylvania 19081, USA}
\author{Nobuyuki Matsumoto}
\affiliation{Department of Physics, University of Tokyo, Bunkyo, Tokyo 113-0033, Japan}
\author{Yoichi Aso}
\affiliation{Department of Physics, University of Tokyo, Bunkyo, Tokyo 113-0033, Japan}
\author{Masaki Ando}
\affiliation{Department of Physics, University of Tokyo, Bunkyo, Tokyo 113-0033, Japan}
\affiliation{National Astronomical Observatory of Japan, Mitaka, Tokyo 181-8588, Japan}

\date{\today}

\begin{abstract}
  An optical ring cavity is used to place the first 
  laboratory constraints on parity-odd
  nonrenormalizable Lorentz violation.
  Variations in resonant frequencies 
  are limited to parts in $10^{15}$.
  Absolute sensitivity to Lorentz-violating
  operators of mass dimension 6
  is improved by a factor of a million over
  existing parity-even microwave-cavity bounds.
  Sensitivity to dimension-8 violations
  is improved by fourteen orders of magnitude.
\end{abstract}

%Special relativity, 03.30.+p
%Frequency, measurement of, 06.30.Ft
%Lorentz invariance, 11.30.Cp
%Resonators, laser, 42.60.Da
%models beyond the standard models, 12.60.-i

\maketitle

Lorentz invariance lies at the foundations of
both the Standard Model of particle physics and General Relativity.
However, attempts to reconcile these two theories
have led to the idea that Lorentz invariance
may only be approximate at attainable energies \cite{ks}.
The prospect of measuring quantum-gravity-induced 
Lorentz violation has spurred a large number of experimental
and theoretical studies in recent decades \cite{datatables}.
These studies include modern versions of the classic
Michelson-Morley experiment \cite{mm} that are based on
electromagnetic resonant cavities
\cite{rings,michimura,microwave,parker,optical,parityeven}.
Here we present a search for Lorentz violation
using an optical ring cavity.
We obtain the first bounds on parity-odd
nonbirefringent nondispersive 
Lorentz-violating operators
of mass dimensions $d=6$ and $8$.
Our results also represent the first optical-frequency
laboratory test of Lorentz violation associated
with operators of dimensions $d=6$ and $8$,
yielding sensitivities that are many orders of magnitude beyond
existing microwave constraints.

Lorentz violation introduces a number
of exotic phenomena in resonant cavities,
such as an unconventional relationship
between frequency and wavelength \cite{km02,km09}.
Since rotations are a part of Lorentz symmetry,
violations also typically introduce orientation dependence,
a key signature in resonant-cavity tests.
Our experiment searches for Lorentz violation
by comparing the resonant frequencies $\nu_\pm$ 
of two counter-propagating directions in a ring cavity
as it is rotated in the laboratory.
Significant Lorentz violation would
lead to modulations in the fractional difference
$\Delta\nu/\nu = (\nu_+-\nu_-)/\nu$
at harmonics of the rotation rate 
$\omega_\rot$.
We would also expect slow modulations
at the sidereal frequency
$\omega_\oplus \simeq 2\pi/(\text{23 hr 56 min})$
due to the Earth's rotation.
So variations in $\Delta\nu/\nu$ at frequencies
$\omega_{mm'} = m\omega_\rot + m'\omega_\oplus$
are a sign of Lorentz violation.

Ring-cavity experiments have recently emerged as
promising parity-odd Lorentz tests \cite{rings,michimura}
that complement existing parity-symmetric experiments
based on microwave cavities \cite{microwave,parker}
and Fabry-P\'erot resonators \cite{optical,parityeven}.
The counter-propagating waves in a ring
are parity mirrors of each other.
As a result, the signal $\Delta\nu/\nu$
changes sign under a parity transformation
and can only depend on parity-odd changes
to the resonant frequencies $\nu_\pm$.
Therefore, our experiment is only sensitive
to Lorentz violations that are odd under parity.

While modulations in $\Delta\nu/\nu$
are a generic signature of Lorentz violation,
a complete theoretical model of
the underlying physics is required to
predict the precise form of the modulations
and how they relate to other experiments.
The general description of Lorentz violation
is provided by the Standard-Model Extension (SME),
which encompasses all known physics and all realistic
violations of Lorentz invariance \cite{sme,km09}.

A Lorentz-violating term in
the Lagrange density of the SME
is constructed from a conventional operator
contracted with a constant tensor coefficient
to form an observer-independent scalar.
In natural units with $\hbar=c=1$,
an unconventional term is partially characterized by
the mass dimension $d$ of the operator.
For example, a dimension-$d$ operator in
the pure photon sector contributes through
a term of the generic form
${\mathcal L}\supset {\mathcal K}^{\alpha_1\alpha_2\ldots\alpha_d}
A_{\alpha_1} \partial_{\alpha_3}\ldots \partial_{\alpha_d}A_{\alpha_2}$,
where $A_\alpha$ is the photon field and
${\mathcal K}^{\alpha_1\alpha_2\ldots\alpha_d}$
are constant coefficients that govern
the size of the Lorentz-violating effects.
The operator, constructed from fields and derivatives
of fields, has mass dimension $d$,
and dimensional analysis shows that the dimension
of the coefficient is $4-d$.
By convention, measurements of 
coefficients are reported
in units of GeV$^{4-d}$.

Restriction to renormalizable 
dimensions $d=3$ and $d=4$ yields
the so-called minimal SME (mSME).
The mSME has been studied extensively,
but nonminimal terms with $d>4$
have received comparatively less attention
due to the large variety and complexity
of the higher-order violations \cite{datatables}.
However, recent theoretical work has established
phenomenology that opens up
the nonminimal sector to experimentation \cite{km09}.
The push to consider nonminimal terms in the SME
is motivated in part by the apparent nonrenormalizability of gravity
and by the possibility that higher-order violations with $d>4$ might dominate.
Theories based on noncommutative spacetime coordinates
provide an example where Lorentz violation emerges
in the form of operators of nonrenormalizable dimension only \cite{ncg}.

Within the photon sector of the SME,
a subset of violations cause vacuum birefringence or dispersion,
which can be tightly constrained through astrophysical observations.
However, a relatively large class of
nonbirefringent nondispersive operators exist
that have no leading-order effect
on light propagating {\it in vacuo}.
These are the so-called camouflage terms,
whose coefficients are denoted
$\cf$ for even dimensions $d\geq 6$.
While the camouflage coefficients cannot
be probed directly in astrophysical tests,
they can be tested in cavity-based experiments.
The goal of this work is to search for unconstrained
parity-odd camouflage coefficients
for dimensions $d=6$ and $d=8$.

Many features of Lorentz violation in electromagnetism
can be understood by examining the structure of the Lagrange density.
Terms with $d=4$ have the same number of derivatives as the usual case,
so the effects of $d=4$ Lorentz violation generally scale with 
photon energy or frequency in the same way as conventional physics.
However, for $d>4$, the additional derivatives imply that the
effects of Lorentz violation typically grow with frequency
by a factor of $\nu^{d-4}$ relative to Lorentz-invariant physics.
This scaling gives optical cavities
an inherent advantage over microwave cavities.
We naively expect an increase in sensitivity 
by a factor of $\sim 10^{4(d-4)}$.

The couplings of fields and derivatives to the coefficient
tensors are also the source of anisotropies,
which we detect as modulations in $\Delta\nu/\nu$.
The effects are similar to those in anisotropic dielectrics,
where the coefficients for Lorentz violation act
as a dielectric tensor filling space.
The additional derivatives in the higher-$d$ terms 
imply added complexity and higher harmonics in the
modulations of $\Delta\nu/\nu$.
For example, the $d=6$ camouflage coefficients
can be shown to introduce monopole, dipole, and quadrupole behavior,
while the $d=8$ coefficients also give hexapole and octupole structure.
Our experiment is only sensitive to parity-odd dipole and hexapole anisotropies.

\begin{figure}
  \begin{center}
    \includegraphics[width=\columnwidth]{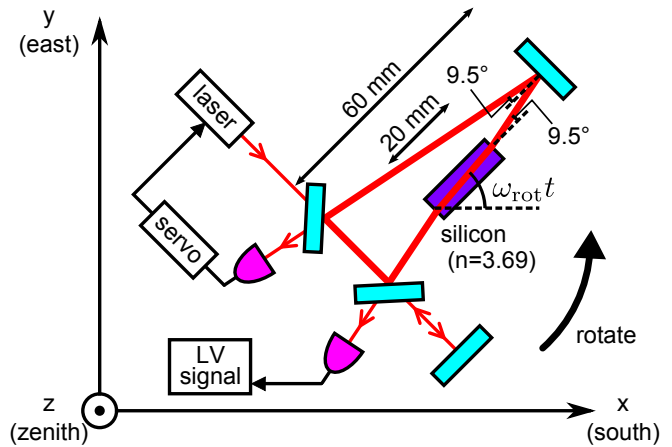}
  \end{center}
  \caption{\label{ringcavity}
    Diagram showing the optical ring cavity
    and its orientation with respect to the
    standard laboratory frame
    defined in Ref.\ \cite{km02}.
    The time is chosen so that $t=0$
    when the beam inside the
    silicon is aligned with the
    $x$ axis (south).
  }
\end{figure}

The experimental setup is illustrated in Fig.\ \ref{ringcavity}.
It is based on a triangular optical ring constructed
from three mirrors rigidly fixed on a spacer.
A silicon piece is placed along one side of the triangle.
The measured refractive index of the silicon piece
was $n=3.69$ at the operational wavelength 
$\lambda = 1550 \unit{nm}$.
All of the optics are placed in a vacuum enclosure
and rotated by a turntable.
A detailed description of the apparatus
can be found in Ref.\ \cite{michimura}.
The silicon provides additional asymmetry
that increases the number of SME coefficients
that can be accessed by our experiment.
In particular, 
calculation shows that without the silicon
we lose sensitivity to all $d=6$ camouflage coefficients.
It also reduces the number of $d=8$ coefficients that can be tested.
The loss in sensitivity stems from the fact
that dipole effects cancel 
around a closed path without matter.

To measure the resonant-frequency difference between
the two counterpropagating directions,
we use a double-pass configuration \cite{doublepass}.
The laser beam is fed into the ring cavity
in the counterclockwise direction.
The frequency of the laser beam is stabilized
to the counterclockwise resonance using
the H\"ansch-Couillaud method \cite{Hansch}.
The transmitted light of the counterclockwise beam
is then reflected back into the cavity in clockwise direction.
This creates the signal that is proportional to
the frequency difference between the counterclockwise beam
and the clockwise resonance.
The double-pass configuration
ensures that most of the effects from environmentally induced
cavity-length fluctuations are canceled
in the differential frequency measurement.

The data used in this analysis were taken at the University of Tokyo
for 393 days between August 2012 and September 2013.
Positive and reverse rotations of $420^{\circ}$
were repeated alternately at the rotation speed
$\omega_\rot=30^{\circ}/\text{sec}$.
Our analysis only uses the middle $360^{\circ}$ of the rotation
where the rotation rate is constant
and fluctuations from the Sagnac effect can be neglected.
The ring cavity was rotated approximately 1.7 million
times during the data acquisition.

To test for Lorentz violation,
one could perform a direct search 
for variations at the frequencies 
$\omega_{mm'} = m\omega_\rot + m'\omega_\oplus$.
However, a demodulation method is appropriate
since $\omega_\rot \gg \omega_\oplus$.
We first consider a decomposition of
the frequency difference into harmonics of $\omega_\rot$,
\begin{equation}
  \frac{\Delta \nu}{\nu} = 
  \sum_{m > 0} \big[ C_m \cos{(m \omega_\rot t)} + S_m \sin{(m \omega_\rot t)} \big] \ .
  \label{demod1}
\end{equation}
A turntable rotation of $180^\circ$ effectively
interchanges the two counterpropagating solutions,
reversing the sign of $\Delta\nu/\nu$.
This implies that we can restrict attention to odd values of $m$.
The dipole anisotropies introduced by the $d=6$
SME coefficients can only give rise to $m=1$.
The dipole and hexapole structure for $d=8$ can give $m=1$ and $m=3$.
We therefore restrict attention to $m=1,3$ harmonics.
Note, however, that Lorentz-violating operators
of higher dimension can lead to harmonics with $m\geq 5$.

The amplitudes $C_m$ and $S_m$ vary at harmonics 
of the sidereal frequency $\omegao$ and can be expanded as
\begin{align}
  C_m &= \sum_{\mpr \ge 0} \big[ \CCmm \cos{(\mpr\alpha)} + \CSmm \sin{(\mpr\alpha)} \big]\ ,
  \nonumber \\ 
  S_m &= \sum_{\mpr \ge 0} \big[ \SCmm \cos{(\mpr\alpha)} + \SSmm \sin{(\mpr\alpha)} \big]\ ,
  \label{demod2}
\end{align}
where $\alpha = \omegao \To$ is the right ascension of the local zenith\ \cite{km09}.
Any nonnegative $m'$ can contribute,
but the multipole structure predicted by the SME
gives $m' = 0,1$ for $d=6$ and $m'=0,1,2,3$ for $d=8$,
so we limit our focus to $0\leq m'\leq 3$.

\begin{figure}
\begin{center}
\includegraphics[width=\columnwidth]{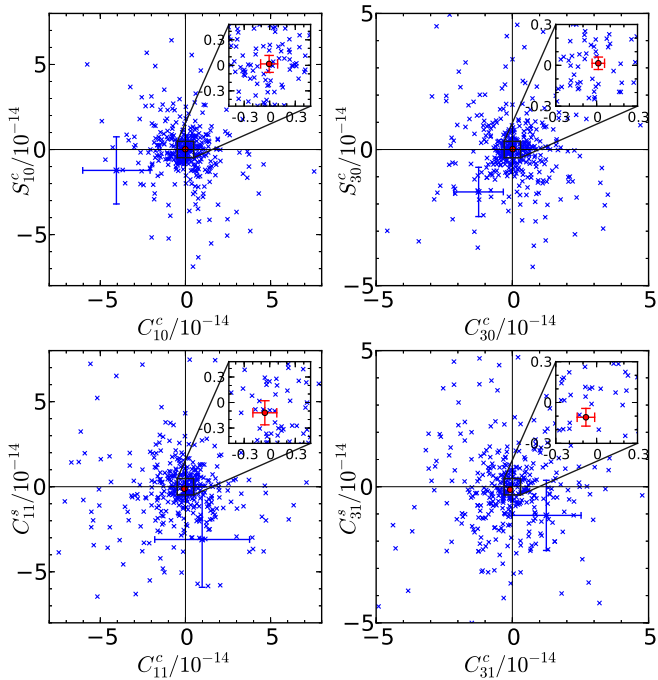}
\end{center}
\caption{\label{phasespace} 
  Examples of the modulation-amplitude measurements.
  The 393 one-day values of $\CCmm$, $\CSmm$, $\SCmm$, and $\SSmm$ are
  shown with $\times$ symbols.
  Error bars are omitted except for one representative point.
  The mean of the 393 points is shown as a red dot with error bars.}
\end{figure}

\begin{table}
  \renewcommand{\arraystretch}{1.1}
  \begin{tabular}{c|c||c|c}
    Amplitude & Measurement & Amplitude & Measurement \\
    \hline\hline
    $\CC{10}$ & $-0.1\pm1.0$ & $\CC{30}$ & $0.12\pm0.46$ \\ 
    $\SC{10}$ & $ 0.2\pm1.0$ & $\SC{30}$ & $0.15\pm0.46$ \\
    $\CC{11}$ & $-0.6\pm1.4$ & $\CC{31}$ & $-0.79\pm0.64$ \\
    $\CS{11}$ & $-1.2\pm1.4$ & $\CS{31}$ & $-1.1\pm0.65$ \\ 
    $\SC{11}$ & $-0.3\pm1.4$ & $\SC{31}$ & $-0.48\pm0.64$ \\ 
    $\SS{11}$ & $ 1.0\pm1.4$ & $\SS{31}$ & $-0.51\pm0.65$ \\ 
    $\CC{12}$ & $-0.9\pm1.4$ & $\CC{32}$ & $-1.1\pm0.65$ \\ 
    $\CS{12}$ & $-0.2\pm1.4$ & $\CS{32}$ & $ 0.57\pm0.65$ \\ 
    $\SC{12}$ & $-0.1\pm1.4$ & $\SC{32}$ & $-0.46\pm0.65$ \\ 
    $\SS{12}$ & $ 1.0\pm1.4$ & $\SS{32}$ & $ 0.21\pm0.65$ \\ 
    $\CC{13}$ & $-0.8\pm1.4$ & $\CC{33}$ & $ 0.40\pm0.65$ \\ 
    $\CS{13}$ & $ 0.2\pm1.4$ & $\CS{33}$ & $ 0.16\pm0.65$ \\ 
    $\SC{13}$ & $-0.5\pm1.4$ & $\SC{33}$ & $-0.36\pm0.64$ \\ 
    $\SS{13}$ & $ 0.6\pm1.4$ & $\SS{33}$ & $ 0.75\pm0.65$ 
  \end{tabular}
  \caption{\label{amplitudes} 
    Constraints on the twenty-eight $\Delta\nu/\nu$ modulation
    amplitudes for $m=1,3$ and $m'=0,1,2,3$.
    All values are in units of $10^{-15}$. }
\end{table}

The analysis starts by demodulating the data
at frequencies $\omega_\rot$ and $3 \omega_{\rot}$
to extract the amplitudes $C_m$ and $S_m$ in Eq.\ \rf{demod1}
for each rotation.
Time series data for $C_m$ and $S_m$ are split into
one-day intervals and fit to Eq.\ \rf{demod2}
by the least-squares method to extract the modulation amplitudes
$\CCmm$, $\CSmm$, $\SCmm$, and $\SSmm$ for each day.
Data sets for some of the amplitudes are shown in Fig.\ \ref{phasespace},
as illustrations. Taking the weighted average over the 393 days
gives our measured values for the modulation amplitudes,
which are listed in Table \ref{amplitudes}.

The error bars for $m=1$ amplitudes are approximately two times larger 
than the errors for $m=3$ amplitudes. The measured noise level in relative 
frequency was $5 \times 10^{-13} \unit{/\sqrt{Hz}}$ at the rotation 
frequency $\omega_{\rot}$ and $4 \times 10^{-13} \unit{/\sqrt{Hz}}$ 
at $3 \omega_{\rot}$ when the turntable is not rotated. However, 
error bars in Table \ref{amplitudes} are more than one order of magnitude 
larger than the expected value from these noise levels. This suggests 
the effect of turntable vibration, which is likely to be larger at the 
rotation frequency. The exact vibration level was not measurable in the 
current setup since the rotation is reciprocated during the data acquisition.

A number of systematic effects was studied. 
A major source of the systematic offset was the tilt of the turntable,
which creates a slight change in the alignment of the incident beam.
This effect was estimated to be less than 10\% of the statistical error.
The Sagnac effect also gives systematic offset to the modulation amplitudes
if the rotational speed of the turntable fluctuates in $1/m$ of a rotational
period. The measured fluctuation was less than $1 \unit{mrad/sec}$,
and this effect is more than 4 orders of magnitude below our statistical error.
Also, there was 3\% uncertainty in the calibration of the signal for Lorentz violation.
This uncertainty originated mainly from a slight drift of
a nonzero detuning in the laser frequency servo.

To get constraints on SME coefficients,
we consider each dimension $d=6$ and $d=8$ separately
and place constraints under the assumption that
only one of the two sets of coefficients is nonzero.
The calculation outlined in Ref.\ \cite{mewes}
yields the relationship between the SME camouflage coefficients $\cf$
and the modulation amplitudes $\CCmm$, $\CSmm$, $\SCmm$, and $\SSmm$.
Each of the amplitudes is a linear combination of
SME coefficients that depends on
the orientation, length, and index of refraction
of each arm of the cavity
and  the colatitude ($\chi=54.3^{\circ}$) of the laboratory.

For $d=6$, variations in the frequency difference
with $m=1$ and $m'=0,1$ arise through
linear combinations of three camouflage coefficients.
The measurements of the modulation amplitudes yield
individual constraints on these three coefficients.
The results are summarized in Table \ref{results}.
Note that SME coefficients $\cf$ with $m\neq 0$ are complex,
so we bound both the real and imaginary pieces.
There are a total of three parity-odd camouflage coefficients for $d=6$,
so our experiment constrains the entire
coefficient space accessible to parity-odd cavity experiments.

\begin{table}
  \renewcommand{\arraystretch}{1.5}
  \begin{tabular}{c|c}
    Coefficient & Measurement \\
    \hline\hline
    $\cfdnjm{6}{110}$ & $(-0.1\pm1.5) \times 10^{3} \unit{GeV^{-2}}$ \\
    $\Recfdnjm{6}{111}$ & $(0.8\pm1.1) \times 10^{3} \unit{GeV^{-2}}$ \\
    $\Imcfdnjm{6}{111}$ & $(-0.6\pm1.0) \times 10^{3} \unit{GeV^{-2}}$ \\
    \hline
    $\cfdnjm{8}{310} - 0.020\cfdnjm{8}{110}$ & $(-0.2\pm1.9) \times 10^{19} \unit{GeV^{-4}}$ \\
    $\Re\big[\cfdnjm{8}{311} - 0.020\cfdnjm{8}{111}\big]$ & $(1.4\pm1.3) \times 10^{19} \unit{GeV^{-4}}$ \\
    $\Im\big[\cfdnjm{8}{311} - 0.020\cfdnjm{8}{111}\big]$ & $(0.1\pm1.3) \times 10^{19} \unit{GeV^{-4}}$ \\
    $\cfdnjm{8}{330}$ & $(-0.8\pm3.3) \times 10^{19} \unit{GeV^{-4}}$ \\
    $\Recfdnjm{8}{331}$ & $(-0.3\pm1.9) \times 10^{19} \unit{GeV^{-4}}$ \\
    $\Imcfdnjm{8}{331}$ & $(-2.8\pm1.9) \times 10^{19} \unit{GeV^{-4}}$ \\
    $\Recfdnjm{8}{332}$ & $(2.2\pm1.3) \times 10^{19} \unit{GeV^{-4}}$ \\
    $\Imcfdnjm{8}{332}$ & $(0.2\pm1.3) \times 10^{19} \unit{GeV^{-4}}$ \\
    $\Recfdnjm{8}{333}$ & $(-0.1\pm1.6) \times 10^{19} \unit{GeV^{-4}}$ \\
    $\Imcfdnjm{8}{333}$ & $(-0.1\pm1.6) \times 10^{19} \unit{GeV^{-4}}$ 
  \end{tabular}
  \caption{\label{results} 
    Measurements of SME $d=6,8$ camouflage coefficients
    with $1\sigma$ errors.}
\end{table}

Considering only $d=8$ coefficients for Lorentz violation,
we find that ten combinations
of coefficients contribute to modulations
of the frequency difference $\Delta\nu/\nu$.
Consequently, the bounds on the twenty-eight amplitudes in Table \ref{amplitudes}
reduce to ten bounds on combinations of $d=8$ SME coefficients.
These bounds are also summarized in Table \ref{results}.
There are a total of thirteen parity-odd camouflage coefficients for $d=8$,
so three linear combinations of coefficients remain untested.
These may be accessed by future ring-cavity experiments with
different configurations, yielding sensitivities to different
combinations of coefficients. 

The results in Table \ref{results} are the first
bounds on parity-odd camouflage coefficients for Lorentz violation.
The current best bounds on the parity-even coefficients
come from the microwave-cavity experiment in Ref.\ \cite{parker}.
While this experiment and our experiment
probe two independent sets of Lorentz violations,
a comparison of the two illustrates the improvement
in sensitivity that results from higher frequencies.
Naive estimates suggest improvements of roughly
eight orders of magnitude for $d=6$ and
sixteen orders of magnitude for $d=8$ may be possible.
However, due to noise differences,
our experiment achieves an increase in sensitivity
that is closer to a factor of a million for $d=6$
and a factor of $10^{14}$ for $d=8$.

Our bounds on SME coefficients
for nonrenormalizable dimensions $d=6$ and $d=8$
are consistent with zero at $2\sigma$,
showing no significant evidence for Lorentz violation.
We achieved raw sensitivity on the order of $10^{-15}$ 
to orientation dependence in the
frequency difference $\Delta\nu/\nu$.
Parity-even optical cavities have achieved
sensitivities at the $10^{-17}$ level in 
tests of the mSME \cite{parityeven},
suggesting the potential for a hundredfold improvement
in future cavity tests of higher-order Lorentz violation.

\acknowledgements
We thank Alan Kosteleck\'y and
Shigemi Otsuka for useful discussions.
This work was supported by Grant-in-Aid
for JSPS Fellows No. 25$\cdot$10386.

\end{document}